\documentclass[sigconf,natbib=true]{acmart}

\makeatletter
\def\@ACM@checkaffil{
    \if@ACM@instpresent\else
    \ClassWarningNoLine{\@classname}{No institution present for an affiliation}%
    \fi
    \if@ACM@citypresent\else
    \ClassWarningNoLine{\@classname}{No city present for an affiliation}%
    \fi
    \if@ACM@countrypresent\else
        \ClassWarningNoLine{\@classname}{No country present for an affiliation}%
    \fi
}
\makeatother

\usepackage{booktabs} 
\usepackage{subfigure}
\usepackage{bm}
\usepackage{xspace}
\usepackage{multirow}
\usepackage{footmisc}

\usepackage{amssymb}
\usepackage{array}
\usepackage{graphicx}
\usepackage{clrscode}
\usepackage{balance}
\usepackage{subfigure}
\usepackage{multirow}
\usepackage{multicol}
\usepackage{float}
\usepackage{color}
\usepackage{xcolor}
\usepackage{amsopn}
\usepackage{mathrsfs}
\usepackage{mathtools}
\usepackage{amsmath}
\usepackage{booktabs}
\usepackage{arydshln}
\usepackage{hyperref}
\usepackage{blkarray}
\usepackage{enumerate}
\usepackage{courier}
\usepackage{mathrsfs}
\usepackage{rotating}
\usepackage{bm}
\usepackage{subfigure}
\usepackage{array}
\usepackage{ragged2e}
\usepackage{amsthm}
\usepackage{enumitem}
\usepackage{marvosym}

\usepackage{bbding}
\usepackage[ruled,linesnumbered]{algorithm2e}

\newcommand{\paratitle}[1]{\vspace{1.5ex}\noindent\textbf{#1}}
\newcommand{\ie}{\emph{i.e.,}\xspace}

\newcommand{\eg}{\emph{e.g.,}\xspace}

\newcommand{\ignore}[1]{}

\AtBeginDocument{%
  \providecommand\BibTeX{{%
    \normalfont B\kern-0.5em{\scshape i\kern-0.25em b}\kern-0.8em\TeX}}}

\setcopyright{acmcopyright}
\copyrightyear{2022}
\acmYear{2022}
\acmDOI{xx.xxxx/xxxxxxx.xxxxxxx}

\acmConference[Conference acronym 'XX]{Make sure to enter the correct
  conference title from your rights confirmation emai}{June 03--05,
  2018}{Woodstock, NY}
\acmBooktitle{Woodstock '18: ACM Symposium on Neural Gaze Detection, June 03--05, 2018, Woodstock, NY}
\acmPrice{15.00}
\acmISBN{978-1-4503-XXXX-X/18/06}

\begin{document}

\title[Multi-grained Hypergraph Interest Modeling for Conversational Recommendation]{\texorpdfstring{Multi-grained Hypergraph Interest Modeling for \\ Conversational Recommendation}{Multi-grained Hypergraph Interest Modeling for Conversational Recommendation}}

\author{Chenzhan Shang}
\email{czshang@outlook.com}
\affiliation{
    \institution{School of Information}
    \institution{Renmin University of China}
}

\author{Yupeng Hou}
\email{houyupeng@ruc.edu.cn}
\affiliation{
    \institution{Gaoling School of Artificial Intelligence}
    \institution{Renmin University of China}
}

\author{Wayne Xin Zhao$^{\dagger}$\textsuperscript{\Letter}}
\email{batmanfly@gmail.com}
\affiliation{
    \institution{Gaoling School of Artificial Intelligence}
    \institution{Renmin University of China}
}
\thanks{$\dagger$ Beijing Key Laboratory of Big Data Management and Analysis Methods.}
\thanks{\Letter\ Corresponding author.}

\author{Yaliang Li}
\email{yaliang.li@alibaba-inc.com}
\affiliation{
    \institution{Alibaba Group}
}

\author{Jing Zhang}
\email{zhang-jing@ruc.edu.cn}
\affiliation{
    \institution{School of Information}
    \institution{Renmin University of China}
}

\begin{abstract}
Conversational recommender system~(CRS) interacts with users through multi-turn dialogues in natural language, which aims to provide high-quality recommendations for user's instant information need. 
Although great efforts have been made to develop effective CRS, most of them still focus on the contextual information from the current dialogue, usually suffering from the data scarcity issue. 
Therefore, we consider leveraging historical dialogue data to enrich the limited contexts of the current dialogue session.

In this paper, we propose a novel multi-grained hypergraph interest modeling approach to capture user interest beneath intricate historical data from different perspectives.
As the core idea, we employ \emph{hypergraph} to  represent complicated semantic relations underlying historical dialogues. 
In our approach, we first employ the hypergraph structure to model users' historical dialogue sessions and form a \textit{session-based hypergraph}, which captures \emph{coarse-grained,  session-level} relations.
Second, to alleviate the issue of data scarcity, we use an external knowledge graph and construct a \textit{knowledge-based hypergraph} considering \emph{fine-grained, entity-level} semantics.
We further conduct  multi-grained hypergraph convolution on the two kinds of hypergraphs, and utilize the enhanced representations to develop interest-aware CRS. 
Extensive experiments on two benchmarks \textsc{ReDial} and \textsc{TG-ReDial} validate the effectiveness of our approach on both recommendation and conversation tasks. Code is available at: \textcolor{blue}{\url{https://github.com/RUCAIBox/MHIM}}.

\end{abstract}



\maketitle
\section{INTRODUCTION}

Conversational Recommender System~(CRS) has become a trending research topic in recent years, with the goal of capturing users' instant preferences through multi-turn dialogues  and providing high-quality recommendations.
Compared to traditional recommender systems, CRS can leverage explicit feedback signals from natural language conversations in an interactive way for more precise modeling of user preferences.

In terms of methodology, a typical CRS consists of a conversation module and a recommendation module. The conversation module aims to understand user utterances and generate informative responses. The recommendation module focuses on capturing user preferences from textual signals and recommending appropriate items. 
Several methods~\cite{ear, cpr, saur} propose to utilize item attributes to progressively narrow down candidate item set and generate responses using pre-defined templates. Other studies~\cite{redial, kbrd} construct end-to-end frameworks for both recommending and generating human-like responses. To further improve the performance, researchers integrate multi-type external data~\cite{kgsf, revcore, c2crs} and devise a more controllable conversation module~\cite{ntrd}.

Though great efforts have been made to develop effective CRSs, most of them focus on utilizing the limited contextual information from the \textit{current dialogue session}~(\ie the ongoing conversation), usually suffering from the data sparsity problem.
To address this issue, our solution is inspired by the observation that a user probably has engaged with the CRS several times before the current conversation.
These \textit{historical dialogue sessions} contain crucial evidence for capturing the preference of a user, which are easy to collect in a practical system.
Therefore, we consider comprehensively capturing  user interest that lies beneath intricate historical data to enrich the limited contexts of the current dialogue session.

However, it is non-trivial to leverage historical dialogues for improving CRS, and there still exist  two major challenges.  
First, the intra- and inter-session correlations  among historical dialogues are complicated, where each dialogue that a user invokes is called a \textit{session}. 
A session typically concentrates on a specific topic, involving a small number of important entities (\eg actor or director for movie recommendation). 
Due to the limited dialogue context, it is difficult to accurately capture the 
relatedness among intra-session entities and establish inter-session relations.  
Second, historical data remains scarce in real-world conversational recommendation scenarios. 
The number of users' historical dialogue sessions and items' interaction records both follow a long-tail distribution. Specifically, most of the users only interact with the system a few times, and thus the historical dialogues may not be sufficient for accurate user modeling. In addition, most of the items probably only appear a few times, resulting in difficulty of learning informative entity representations for recommendation.

In light of these challenges, and inspired by the work about hypergraph learning~\cite{hgnn, hypergcn} and history-aware dialogue system~\cite{ham, gupta2021role}, our core idea is to employ \textit{hypergraph}s to represent complicated semantic relations among historical dialogue sessions. 
Different from standard graph, in a hypergraph,  a hyperedge connects more than two vertices~\cite{higher}, which is particularly suitable to model the interrelations among multiple objects (\eg entities).   When applied to our setting,  hypergraph can better represent a dialogue by directly associating multiple  involving entities in an explicit way. 
Besides, different hyperedges also share common vertices, which may be useful to model  
vital attributes for reflecting intrinsic  user interest. 

To this end, in this paper, we propose a novel \textbf{M}ulti-grained \textbf{H}ypergraph \textbf{I}nterest \textbf{M}odeling approach for conversational recommendation, named as \textbf{MHIM}. We consider two major ways to construct hypergraphs for improving CRS. 
First, to capture complicated relations in historical data, we model each session as a hyperedge of items and construct a \textit{session-based hypergraph}.
Second, to alleviate data scarcity, we incorporate an external knowledge graph~(KG) and construct a \textit{knowledge-based hypergraph}.
In order to model the above two kinds of hypergraphs, we introduce multi-grained hypergraph convolution to model historical user interest, with the session-based hypergraph capturing  coarse-grained (\emph{session-level}) user preferences beneath historical data, and the knowledge-based hypergraph capturing fine-grained (\emph{entity-level})  user interest on the KG. 
Besides,  to learn informative entity representations, we propose to pre-train the KG encoder by contrastive subgraph discrimination.  
To make accurate recommendations, we devise a hypergraph-aware attention module to derive user representation, which considers multi-grained user preferences learned from historical data.
For the conversation task, historical dialogue sessions are further leveraged to build an interest-aware response generator.
 
To our knowledge, it is the first time that historical data is leveraged to model user interest for CRS via hypergraph modeling.
We firstly construct multi-grained hypergraphs to model user preferences from dialogues, and then propose a hypergraph convolution layer to learn informative entity and user representations for conversational recommendation.
Extensive experiments on two public CRS datasets have demonstrated the effectiveness of our approach in both recommendation and conversation tasks, by comparing a number of competitive baselines.
\section{PRELIMINARIES}

In this section, we first formulate the  task of conversational recommendation, and then introduce the definition of hypergraph.

\paratitle{Task Description}. Conversational recommendation aims to provide high-quality recommendations through a multi-turn dialogue with users. During the dialogue, the system generates responses for clarification or makes relevant recommendations, until the user accepts the results or exits.
Typically, a CRS consists of two major components, namely the recommendation module and the conversation module. These two modules should be integrated seamlessly to fulfill the recommendation goal~\cite{kbrd, kgsf}.
Formally, let $\mathcal{I}$ denote the item set and $\mathcal{U}$ denote the user set. A conversation $\mathcal{C}$ is a sequence composed of utterances (\ie a text sentence) occurring between a CRS and a user, denoted by $\mathcal{C} = \{ S_t \}_{t = 1}^{n}$. Each utterance is composed of a sequence of words. 
Given an $n$-turn conversation, the goal of CRS is to generate responses to the user, including both the recommendation set $\mathcal{I}_{n+1}$ and the reply utterance $S_{n+1}$.

\paratitle{Current and Historical Dialogue Sessions}. Furthermore, in a CRS, a user probably interacts with the system multiple times.
For example, in an e-commerce platform, a user might chat with the intelligent assistant on different days seeking different types of products, reflecting the user's long-term and diverse interests.
Specifically, the historical conversations that a user involved in are called \textit{historical dialogue sessions}, and the ongoing conversation is called \textit{current dialogue session}~\cite{uccr}.
Intuitively, it is useful to consider the historical dialogue sessions when inferring user preference about the items during the current dialogue session.
By leveraging historical data, the system has potential to learn more precise user preference representations for proper recommendations. 

\paratitle{Hypergraph}.
To better capture the semantics of a dialogue session, a key point is to effectively model the complex interrelations of entities mentioned in the session. 
For a traditional graph structure, an edge connects two vertices, while hypergraph generalizes the concept of edge to connect more than two vertices~\cite{higher, hgnn}. A hypergraph is defined as $\widetilde{\mathcal{G}} = (\mathcal{V}, \mathcal{H})$, which includes a vertex set $\mathcal{V}$ and a hyperedge set $\mathcal{H}$. The hypergraph can be represented by an incidence matrix $\mathbf{H} \in \{ 0, 1 \}^{|\mathcal{V}|\times|\mathcal{H}|}$, with each entry $\text{H}_{v, h}$ indicating whether a vertex $v$ is connected by a hyperedge $h$:
\begin{equation}
\text{H}_{v, h} = \left\{
\begin{array}{ll}
    1 & \textrm{if}\ v \in h, \\
    0 & \textrm{if}\ v \notin h.
\end{array}
\right.
\end{equation}
For a vertex $v \in \mathcal{V}$, its degree is defined as $d(v) = \sum_{h\in \mathcal{H}} \text{H}_{v, h}$. For an edge $h \in \mathcal{H}$, its degree is defined as $\delta(h) = \sum_{v \in \mathcal{V}} \text{H}_{v, h}$. Further, let $\mathbf{D} \in \mathbb{N}^{|\mathcal{V}|\times|\mathcal{V}|}$ and $\mathbf{B} \in \mathbb{N}^{|\mathcal{H}|\times|\mathcal{H}|}$ denote the diagonal matrices of the vertex degrees and the edge degrees, respectively.
We propose to model historical data as session-based and knowledge-based hypergraphs.
The former constructs the items in the same dialogue as a hyperedge, and the latter constructs a target  item and its neighborhood on KG as a hyperedge.
As will be shown in Section~\ref{sec-multi-grained}, these hypergraphs can capture coarse-grained and fine-grained user interest, which benefits item recommendation. 
\begin{figure*}
\includegraphics[width=0.95\textwidth]{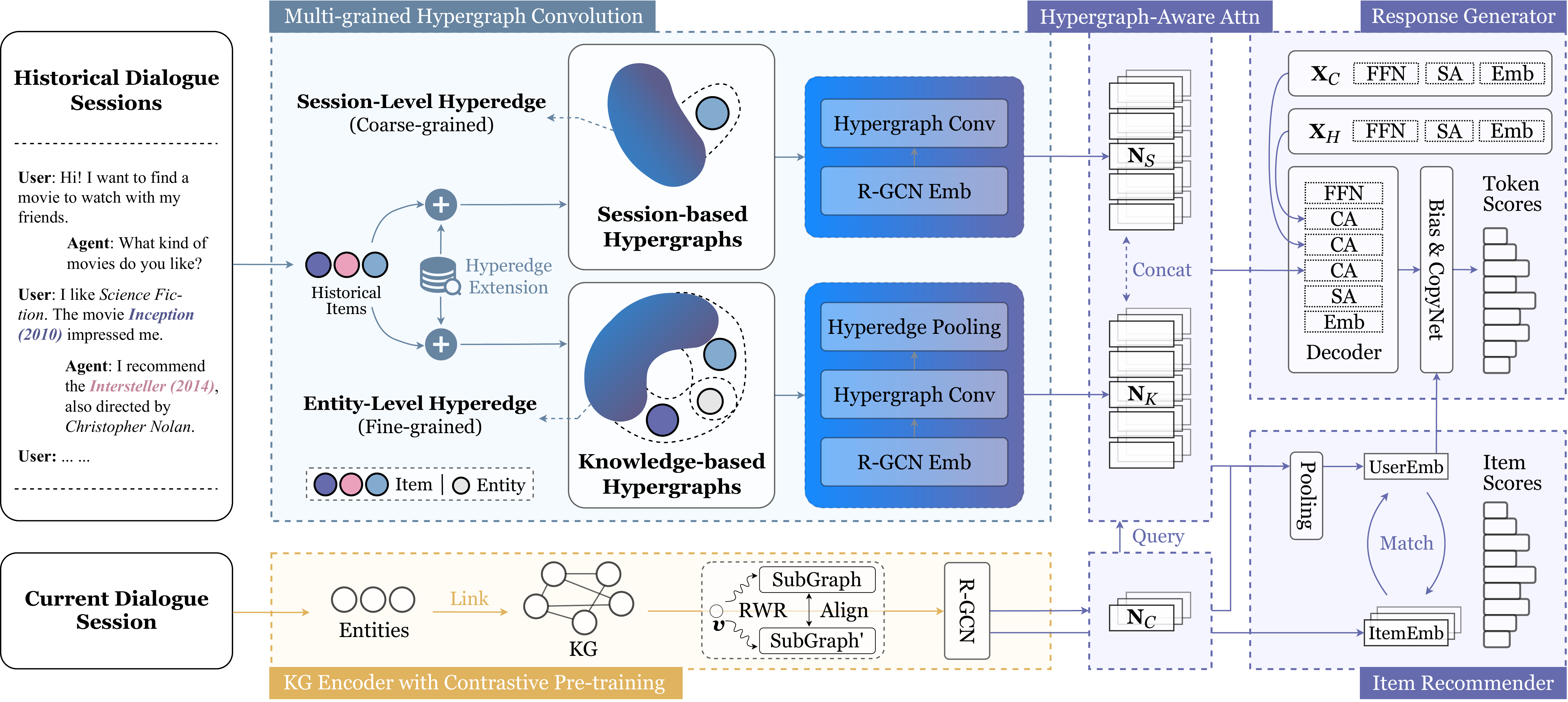}
\caption{The overview of our model in a movie recommendation scenario. We first encode KG and then capture user interest through multi-grained hypergraph convolution. The learned entity representations are further used for item recommendation and response generation. Moreover, ``SA'' and ``CA'' denote self-attention and cross-attention layers, respectively.}
\label{approach}
\end{figure*}

\section{APPROACH}
\label{sec-approach}

In this section, we present the proposed \textbf{\underline{M}}ulti-grained \textbf{\underline{H}}ypergraph \textbf{\underline{I}}nterest \textbf{\underline{M}}odeling approach for conversational recommendation, named as \textbf{MHIM}.
We first introduce how to encode external KG to represent fine-grained entities.
Then, based on the learned entity representations, we present our method for modeling historical user interest through multi-grained hypergraph convolution.
We finally describe our solutions for both recommendation and conversation tasks utilizing the above user interest representations.
The overview illustration of our proposed model is presented in Figure~\ref{approach}.

\subsection{Knowledge Graph Encoding with Contrastive Pre-training}
\label{sec-encoding}

As KG provides important external information for conversational recommendation task, we first present our representation learning method for task-specific KG. Then, we describe the enhanced training method by introducing contrastive subgraph pre-training. 

\subsubsection{Task-specific KG Encoder}
\label{sec-encoder}

Following~\cite{kbrd, kgsf}, to capture the semantics of dialogue contents, we extract the basic semantic units~(\ie \textit{entity}) from utterances and link them to KG entries, and  then extend these entities with two-hop search. The mentioned entities and the extended entities form the \emph{task-related KG} for CRS.  
We adopt the widely-used \textsc{DBpedia}~\cite{dbpedia} and \textsc{CN-DBpedia}~\cite{cn-dbpedia} as external KG. In KG,  a semantic fact is usually denoted as a triplet $\langle e, r, e'\rangle$, where $e, e' \in \mathcal{E}$ are entities and $r \in \mathcal{R}$ is an entity relation.
In our setting, an item is also an entity, which means $\mathcal{I} \subseteq \mathcal{E}$. 
Considering that the relations are crucial for learning entity representations, we utilize R-GCN~\cite{r-gcn} to develop the KG encoder $f_q(\cdot)$, which improves the basic GCN architecture by  explicitly modeling the relational semantics. Through information propagation and aggregation upon KG, we obtain the embedding matrix for entities in $\mathcal{E}$ as $\mathbf{N} \in \mathbb{R}^{|\mathcal{E}| \times d}$.

\subsubsection{Improved KG Encoder by Contrastive Subgraph Discrimination}
\label{sec-contrastive}

However, limited by the size of CRS datasets, it is usually difficult to construct a sufficiently large task-related KG for training the KG encoder $f_q(\cdot)$.  Thus, we consider utilizing the \emph{large-scale, original} KG for improving the KG encoder.
To reduce the influence of irrelevant information, we only keep the relations that occur in the CRS datasets  and utilize them to span  the connected component from the original KG, called \emph{extended KG}. Compared with task-related KG, the extended KG contains more \emph{relevant, diverse} entities, since the contained relations are from the CRS dataset. Our idea is to introduce contrastive pre-training on the extended KG to improve the KG encoder.
Specifically, we propose to leverage subgraphs as contrastive instances and use \textit{subgraph instance discrimination} as our pre-training task. This task treats each subgraph as a distinct class and learns to discriminate between them.

We adopt random walk with restart~\cite{gcc}, in which the walk returns back to starting vertex $v$ with a positive probability.
The subgraph $\widehat{\mathcal{G}}$ is finally induced by the collected vertices during the random walk. To generate the subgraph representation, we first employ the KG encoder (Section~\ref{sec-encoder}) to encode the nodes in a subgraph instance, and then sum and normalize the node embeddings as subgraph representation. 
To construct the contrastive samples, we consider two subgraphs derived from the same starting vertex as a similar instance pair, and others as  dissimilar instance pairs.

Consider an encoded query $\bm{q}$ (\ie the target subgraph) and a dictionary of $M+1$ encoded keys $\{ \bm{k}_0, \ldots, \bm{k}_M \}$ (\ie the contrastive samples), we assume that there is a single key (\ie the positive sample) that $\bm{q}$ matches in the dictionary, denoted by $\bm{k}_+$. We adopt the InfoNCE contrastive loss  in our work:
\begin{equation}
\mathcal{L} = - \log \frac{\exp ( \bm{q}^\top \bm{k}_+ / \tau )}{ \sum_{i=0}^{M} \exp (\bm{q}^\top \bm{k}_i / \tau) },
\end{equation}
where $\tau$ is the temperature hyper-parameter, $\bm{q}$ and $\bm{k}$ are subgraph instance representations encoded by two separate R-GCN encoder $f_q$ and $f_k$, denoted by $\bm{q} = f_q( \widehat{\mathcal{G}}_q )$ and $\bm{k} = f_k( \widehat{\mathcal{G}}_k )$. 
Here, we follow~\cite{moco} to set two different encoders and use a momentum-based update strategy, which maintains a queue of samples from preceding mini-batches. Formally, the parameters of $f_q$ and $f_k$ are denoted as $\mathbf{\Theta}_q$ and $\mathbf{\Theta}_k$, respectively. We update $\mathbf{\Theta}_k$ by $\mathbf{\Theta}_k \leftarrow m \mathbf{\Theta}_k + (1-m)\mathbf{\Theta}_q$, where $m \in [0, 1)$ is the momentum hyper-parameter.

Note that our goal is to pre-train a more capable KG encoder $f_q$, while $f_k$ is introduced to improve the training of $f_q$, which will be discarded after pre-training. 

\subsection{Multi-grained Hypergraph for Historical User Interest Modeling}
\label{sec-multi-grained}

In this subsection, we first introduce the general hypergraph convolution for feature transformation and aggregation.
Then, we propose multi-grained hypergraph convolution for user interest modeling, with the session-based hypergraph aggregating coarse-grained user preferences beneath historical data, and the knowledge-based hypergraph capturing fine-grained user interest on the KG.

\subsubsection{Hypergraph Convolution}

In this paper, we develop a hypergraph convolutional network to capture high-order relations. Following~\cite{hgnn, hgnn-attn}, we define our hypergraph convolution as: 
\begin{equation}
\bm{x}_i^{(l+1)} = \sum_{v=1}^{|\mathcal{V}|}\sum_{h=1}^{|\mathcal{H}|} \text{H}_{i, h} \text{H}_{v, h} \bm{x}_v^{(l)} \mathbf{W}^{(l)},
\end{equation}
where $\bm{x}_i^{(l)}$ is the embedding of the $i$-th vertex in the $l$-th layer, $\mathbf{W}^{(l)} \in \mathbb{R}^{d\times d}$ is the weight matrix between the $l$-th and the $(l+1)$-th layer.
However, the scale of the vertex embeddings may change during training, resulting in numerical instabilities. 
Therefore, we introduce a proper normalization and rewrite the convolution operation in the form of the matrix:
\begin{equation}
\label{eq-hyperconv}
\mathbf{X}^{(l+1)} = \mathbf{D}^{-1} \mathbf{H} \mathbf{B}^{-1} \mathbf{H}^{\top} \mathbf{X}^{(l)} \mathbf{W}^{(l)},
\end{equation}
where $\mathbf{X}^{(l)}$ and $\mathbf{X}^{(l+1)}$ are the input of the $l$-th and $(l+1)$-th layers, respectively.

Firstly, the vertex features are transformed by weight matrix $\mathbf{W}$.
Then, the vertex features are aggregated to form the hyperedge features by transposed incidence matrix $\mathbf{H}^{\top}$, and the related hyperedge features are aggregated to generate refined vertex features by $\mathbf{H}$.
The degree matrices for vertex and hyperedge~(\ie $\mathbf{D}$, $\mathbf{B}$) are introduced for normalization.
Therefore, the hypergraph convolution can be viewed as a two-stage refinement process,  performing ``\emph{vertex-hyperedge-vertex}'' feature transformation upon hypergraph structure.
Note that we do not use nonlinear activation function~(\eg {ReLU}) for hypergraph convolution following~\cite{simplifying, dhcn}.

\begin{figure}
    \centering
    \includegraphics[width=0.47\textwidth]{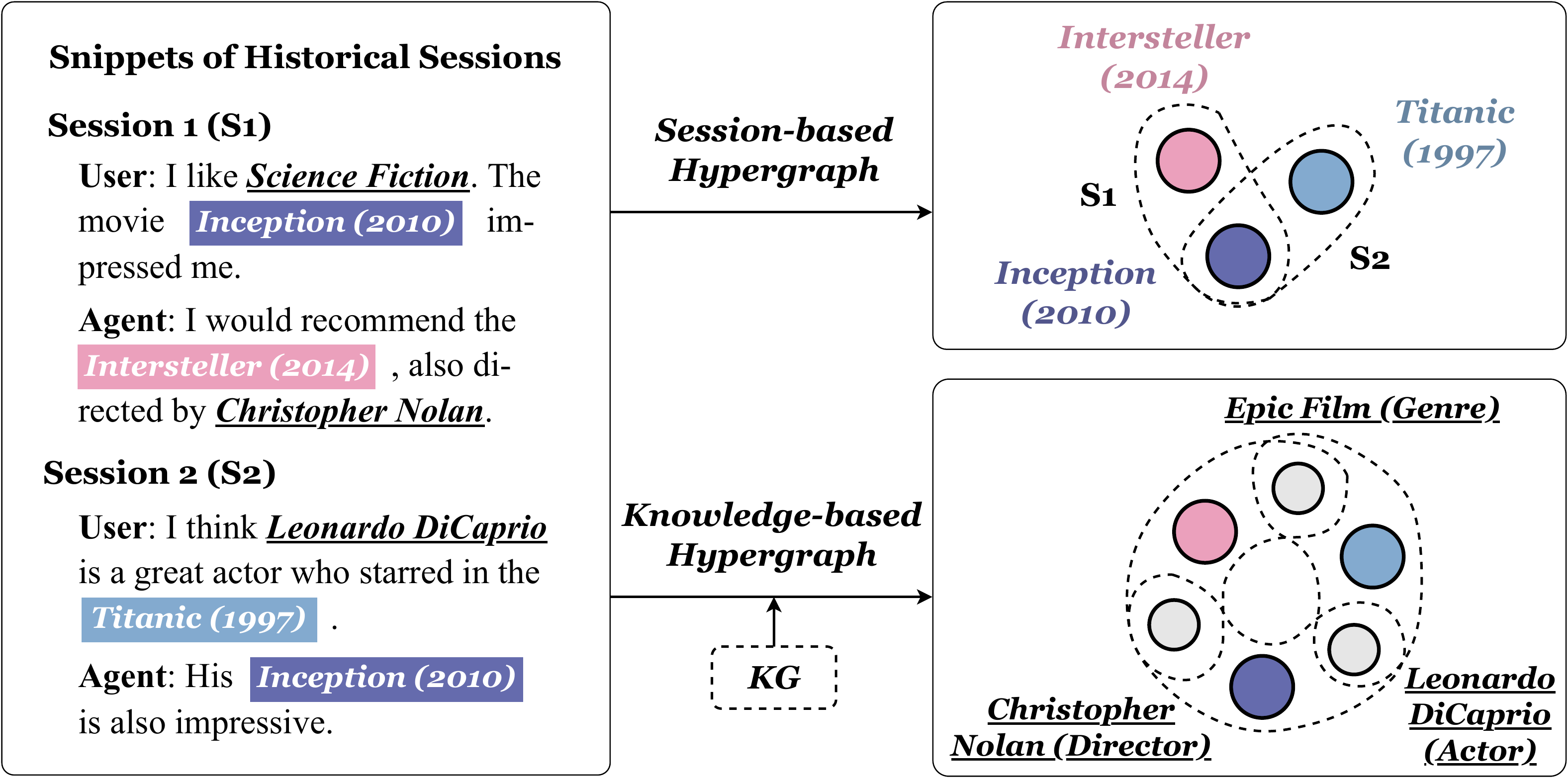}
    \caption{Illustrative examples of how we build session-based and knowledge-based hypergraphs based on the items extracted from historical dialogue sessions.}
    \label{hypergraph}
\end{figure}

\subsubsection{Session-based Hypergraph Convolution}
\label{sec-session-based}

To enrich the limited context of the current dialogue session, we leverage the historical dialogue sessions of a user for preference modeling. 
We observe that each session usually concentrates on a single topic.
For example, when a user is looking for a science fiction, the items that appear  in the conversation utterances would be related to that topic.
In addition, different sessions from one certain user probably share common items, which indicates intrinsic user preference.

Inspired by this observation, we propose to model each historical dialogue session as a hyperedge.
Specifically, the items that appear in the same session are extracted as vertices to build a hyperedge, with the chronological order of the items being ignored.
Then, all the hyperedges corresponding to the user connect with each other via shared items, which constitute a \textit{session-based hypergraph}.
For example, in Figure~\ref{hypergraph}, both sessions contain two movie items, which correspond to the two hyperedges in the session-based hypergraph, respectively. Moreover, the item ``\textit{Inception~(2010)}'' connects the two hyperedges which appear in both sessions.

Formally, the items that appear in the historical dialogue sessions constitute a \textit{historical item set} $\mathcal{I}_H$, and for session-based hypergraph composed of $\mathcal{I}_H$, the incidence matrix is denoted by $\mathbf{H}_S$, the diagonal matrices of the vertex degrees and the edge degrees are denoted as $\mathbf{D}_S$ and $\mathbf{B}_S$, respectively. We extract the representations of $\mathcal{I}_H$ from encoded entity embeddings as $\mathbf{N}_H$, which is fed into the session-based hypergraph convolution:
\begin{equation}
\label{eq-ns}
\mathbf{N}_S = \textbf{HConv} (\mathbf{H}_S, \mathbf{D}_S, \mathbf{B}_S, \mathbf{N}_H),
\end{equation}
where $\textbf{HConv}(\cdot)$ is a hypergraph convolution defined by Equation~\eqref{eq-hyperconv}, and $\mathbf{N}_S$ is the embedding matrix enhanced by session-based hypergraph convolution for historical item set $\mathcal{I}_H$.

After the \emph{vertex-hypergraph-vertex} feature transformation upon the above session-based hypergraph,
where items from the entire session form the hyperedge,
the item representations aggregate session-level semantics from historical dialogue sessions from such a coarse-grained perspective.

\subsubsection{Knowledge-based Hypergraph Convolution}
\label{sec-knowledge-based}

In a real-world scenario, the number of historical dialogues related to the users follows a long-tail distribution~\cite{long-tail}, which indicates  that most of the users only interact with the system a few times. In this case, the contextual information provided by historical dialogue sessions may be insufficient for discovering intrinsic user preferences. Therefore, we propose to explore user interest upon external KG.

Specifically, we model each historical item~(\ie the items from the historical item set $\mathcal{I}_H$) and its $N$-hop neighbors as a hyperedge. The motivation is that the vertex and the extended neighbors probably share common semantic meaning. Then, all of the hyperedges derived from historical items connect with each other via shared entities, which constitute a \textit{knowledge-based hypergraph}. The anchor nodes which connect hyperedges play an important role in aggregating user interest upon KG, since they may be common attributes shared by different historical items.
For example, in Figure~\ref{hypergraph}, the three items that appear in the historical sessions and their neighbors on KG constitute a knowledge-based hypergraph. Each hyperedge contains a specific item and several entities, and the entities are shared between hyperedges.

Given a knowledge-based hypergraph, it includes a vertex set $\mathcal{E}_K$ composed of historical items and their $N$-hop neighbors. The incident matrix is denoted by $\mathbf{H}_K$, and the diagonal matrices of the vertex degrees and the edge degrees are denoted as $\mathbf{D}_K$ and $\mathbf{B}_K$. The representations of $\mathcal{E}_K$ constitute an embedding matrix $\mathbf{N}'_K$, which is subsequently  fed to knowledge-based hypergraph convolution:
\begin{equation}
\label{eq-nk}
\widetilde{\mathbf{N}}_K = \textbf{HConv} (\mathbf{H}_K, \mathbf{D}_K, \mathbf{B}_K, \mathbf{N}'_K),
\end{equation}
where $\widetilde{\mathbf{N}}_K$ is the embedding matrix enhanced by the knowledge-based hypergraph convolution for $\mathcal{E}_K$.
Then, we construct representations of historical items from $\widetilde{\mathbf{N}}_K$, denoted by $\mathbf{N}_K$. Mean pooling is performed on each hyperedge to obtain item representations.

After the message passing upon the above knowledge-based hypergraph,
where a single item and its neighbors on the KG form the hyperedge,
the item representations aggregate similar semantics from triplets stored in the KG, 
which promotes capturing entity-level user interest from historical dialogue sessions from such a fine-grained perspective.

\subsubsection{Hyperedge Extension with Similar Dialogues}
\label{sec-hyperedge-extension}

To further alleviate the scarcity of user historical dialogues, we propose to perform hyperedge extension with similar dialogues based on interactions with common items. The basic idea is that common items in dialogues usually correspond to similar preferences among users. 

Specifically, we extract items from dialogues and build hyperedges, which form a hyperedge collection. Then, based on the common item ratio between the current dialogue and the hyperedges in the hyperedge collection, a certain number of hyperedges are selected for extension. The extended hypergraphs are further used to construct the hypergraph convolution as described before. 
We devise an adaptive method to determine the scale of hyperedge extension. If the size of a hypergraph is relatively small, we tend to extend it with a smaller number of hyperedges. Such a strategy prevents incorporating too much noise, thus leading to more accurate modeling of the current hyperedge. 

Above, we model historical data as session-based and knowledge-based hypergraphs, which provide coarse-grained and fine-grained perspectives for capturing user preferences.
Such a way allows our model to fuse user interest considering multi-grained semantics, and obtain informative representations for recommendation.

\subsection{User Interest-Aware CRS}

In this subsection, we fuse the item representations enhanced by the proposed multi-grained hypergraph convolution (Section~\ref{sec-multi-grained}) to obtain user representation.
Based on this, we further build a user interest-aware CRS, which consists of an item recommender and a response generator.

\subsubsection{User Representation via Hypergraph-Aware Attention}

Since user's historical interests tend to be diverse, historical dialogue sessions  may not be  related to the current user interest. For example, a user has watched science fictions, comedies and cartoons in the past, but now only wants to watch a science fiction. Our goal is to leverage the related historical interest for enhancing the modeling of current user interest. For this purpose, we propose a hypergraph-aware multi-head attention layer to integrate historical item representations with current entity representations.

Recall that we have learned session-based and knowledge-based item representations, \ie  $\mathbf{N}_{S}$~(Equation~\eqref{eq-ns}) and $\mathbf{N}_{K}$~(Equation~\eqref{eq-nk}), respectively, based on the hypergraph structure. We next take the entity representations from the current dialogue session denoted by $\mathbf{N}_C$  as the query to attend to both 
the item representations in $\mathbf{N}_{S}$ and $\mathbf{N}_{K}$. 
Formally, we calculate the integrated representations of historical items as:
\begin{equation}
\label{eq-nsk}
\mathbf{N}_{SK} = \text{MHA}(\mathbf{N}_C, [\mathbf{N}_S;\mathbf{N}_K], [\mathbf{N}_S;\mathbf{N}_K]),
\end{equation}
where $\text{MHA}(\mathbf{Q}, \mathbf{K}, \mathbf{V})$ defines a multi-head attention function which takes a query matrix $\mathbf{Q}$, a key matrix $\mathbf{K}$ and a value matrix $\mathbf{V}$ as input, following~\cite{transformer}. 

Such a way can leverage related entity tastes from the constructed hypergraph structures, considering both multi-grained (\emph{session-} and \emph{entity-level}) semantics. 
Thus, the current user interest can be enhanced by fusing related historical preferences.  
Then, we adopt a pooling layer~(\eg mean pooling) to fuse historical and current user interests, and obtain the final user representation $\bm{u}$:
\begin{equation}
\bm{u} = \text{Pooling}([\text{Pooling}(\mathbf{N}_{SK});\mathbf{N}_C]).
\end{equation}

\subsubsection{Item Recommendation}

Above, we have obtained the user representation enhanced by multi-grained hypergraph convolution, considering both session-level and entity-level semantics from historical dialogue sessions.
We further calculate the probabilities that recommend the items to user $u$: 
\begin{equation}
P_{rec} = \text{Softmax}(\bm{u}\cdot\mathbf{N}_I^\top),
\end{equation}
where $\mathbf{N}_I$ denotes the embeddings of all the candidate items from item set $\mathcal{I}$, which is encoded with pre-training following Section~\ref{sec-encoding}.

To learn the parameters of the model, we adopt cross-entropy loss as the objective function:
\begin{equation}
\mathcal{L}_{rec} = - \sum_{j=1}^{B} \sum_{i=1}^{|\mathcal{I}|} [ -(1 - y_{ij})\cdot \log(1 - P_{rec}^{(j)}(i)) + y_{ij} \cdot \log(P_{rec}^{(j)}(i)) ],
\end{equation}
where $B$ is the size of mini-batch, $y_{ij} \in \{ 0, 1 \}$ is the target label.

\subsubsection{Response Generation}

Following prior works~\cite{kbrd, kgsf}, we adopt Transformer~\cite{transformer} to develop an encoder-decoder framework for conversation task. We introduce two separate encoders to encode historical and current dialogues, respectively. Then, the learned representations are fed into the decoder as cross-attention signals.

Formally, in each decoder layer, we first obtain text representations after self-attention and cross-attention layers:
\begin{eqnarray}
\mathbf{A}_{0}^{n} &=& \text{MHA}(\mathbf{R}^{n-1}, \mathbf{R}^{n-1}, \mathbf{R}^{n-1}), \\
\mathbf{A}_{1}^{n} &=& \text{MHA}(\mathbf{A}_{0}^{n}, \mathbf{N}_{SK}, \mathbf{N}_{SK}),
\label{eq-mha-nsk}
\end{eqnarray}
where $\mathbf{R}^{n-1}$ is the embedding matrix from the decoder at $(n-1)$-th layer, and $\mathbf{N}_{SK}$ is the item representations of the current  dialogue enhanced by multi-grained hypergraph convolution (Equation~\eqref{eq-nsk}). Then, we fuse the representations of historical and current dialogue sessions into decoder.
To avoid overfitting on historical dialogues, we introduce a hyper-parameter $\beta$ to achieve the trade-off between these two types of signals:
\begin{eqnarray}
\mathbf{A}_{2}^{n} &=& \text{MHA}(\mathbf{A}_{1}^{n}, \mathbf{X}_C, \mathbf{X}_C), \\
\mathbf{A}_{3}^{n} &=& \text{MHA}(\mathbf{A}_{1}^{n}, \mathbf{X}_H, \mathbf{X}_H), \\
\mathbf{A}_{4}^{n} &=& \beta \cdot \mathbf{A}_{2}^{n} + (1 - \beta) \cdot \mathbf{A}_{3}^{n},
\end{eqnarray}
where $\mathbf{X}_C$ is the embedding matrix output by the current dialogue session encoder, and $\mathbf{X}_H$ is the embedding matrix output by the historical dialogue session encoder.
Note that for the conversation module, the main difference between our method and KGSF~\cite{kgsf} is that we adopt two separate encoders to encode the current and the historical dialogues respectively, and fuse the item representations enhanced by multi-grained hypergraph convolution into the decoder.
Finally, we obtain decoder layer output through a feed-forward network layer following~\cite{transformer}:
\begin{equation}
\mathbf{R}^{n} = \text{FFN}(\mathbf{A}_{4}^{n}).
\end{equation}

Furthermore, the generated responses are expected to reflect user interest and contain diverse recommended items. Therefore, we adopt a user interest-aware bias and another item-related bias generated by the copy mechanism. Given the predicted sequence $y_1, \ldots, y_{i-1}$, the next token probability is calculated as:
\begin{equation}
P_{gen}(y_i|y_1, \ldots, y_{i-1}) = P_1(y_i|\mathbf{R}_i) + P_2(y_i|\bm{u}) + P_3(y_i|\mathbf{R}_i, \bm{u}),
\label{eq-bias}
\end{equation}
where $P_1(\cdot)$ is the vocabulary probability generated by taking the decoder output $\mathbf{R}_i$ as input, $P_2(\cdot)$ is the vocabulary bias generated by user representation $\bm{u}$, and $P_3(\cdot)$ denotes the copy probability where the scores of non-item vocabularies are set to $0$. Finally, the conversation module is trained with the cross-entropy loss:
\begin{equation}
\mathcal{L}_{gen} = - \sum_{i=1}^{B}\sum_{t=1}^{T} \log(P_{gen}(y_t|y_1, \ldots, y_{t - 1})),
\end{equation}
where $B$ is the batch size, and $T$ is the truncated length of utterances.

\section{EXPERIMENT}

To verify the effectiveness of the proposed method \textbf{MHIM}, we conduct extensive experiments and provide detailed analysis.

\begin{table}[t]
    \centering
    \caption{Statistics of the datasets in our experiments.}
    \small
    \label{datasets}
    \begin{tabular}{lrrrr}
        \toprule
        \textbf{Dataset} & \textbf{\#Dialogues} & \textbf{\#Users} & \textbf{\#Items} & \textbf{Sparsity} \\
        \midrule
        ReDial~\cite{redial} & 11,348 & 956 & 6,924 & 99.9843\% \\
        TG-ReDial~\cite{tgredial} & 10,000 & 1,482 & 33,834 & 99.9973\% \\
        \bottomrule
    \end{tabular}
\end{table}

\begin{table*}
\setlength{\abovecaptionskip}{12pt}   
\setlength{\belowcaptionskip}{2pt}
\centering
  \caption{Experimental results on recommendation task. * indicates statistically significant improvement ($p<0.05$) over all baselines. Source refers to the dataset or external knowledge related to the method, where \textit{D} refers to datasets \textsc{ReDial} and \textsc{TG-ReDial}, \textit{E} and \textit{W} refer to the entity-level KG~(\ie\textsc{DBpedia} and \textsc{CN-DBpedia}) and word-level KG~(\ie\textsc{ConceptNet} and \textsc{HowNet}), respectively.
  We abbreviate Recall@$\mathbf{\emph{K}}$, MRR@$\mathbf{\emph{K}}$ and NDCG@$\mathbf{\emph{K}}$ as R@$\mathbf{\emph{K}}$, M@$\mathbf{\emph{K}}$ and N@$\mathbf{\emph{K}}$, respectively.
  }
\setlength{\tabcolsep}{0.8mm}{

\begin{tabular}{llllllllllllll}
\toprule 
\multirow{2}{*}{\textbf{Source}} 
& \multirow{2}{*}{\textbf{Model}} 
& \multicolumn{6}{c} { \textsc{ReDial} } & \multicolumn{6}{c} { \textsc{TG-ReDial} } \\
\cmidrule(lr){3-8} \cmidrule(lr){9-14}
& & \multicolumn{1}{c}{R@10} & \multicolumn{1}{c}{R@50} & \multicolumn{1}{c}{M@10} & \multicolumn{1}{c}{M@50} & \multicolumn{1}{c}{N@10} & \multicolumn{1}{c}{N@50} & \multicolumn{1}{c}{R@10} & \multicolumn{1}{c}{R@50} & \multicolumn{1}{c}{M@10} & \multicolumn{1}{c}{M@50} & \multicolumn{1}{c}{N@10} & \multicolumn{1}{c}{N@50}\\
\midrule 

D & TextCNN~\cite{textcnn} & 0.0644 & 0.1821 & 0.0235 & 0.0285 & 0.0328 & 0.0580 & 0.0097 & 0.0208 & 0.0040 & 0.0045 & 0.0053 & 0.0077 \\ 

D & SASRec~\cite{sasrec} & 0.1117 & 0.2329 & 0.0540 & 0.0593 & 0.0674 & 0.0936 & 0.0043 & 0.0178 & 0.0011 & 0.0017 & 0.0019 & 0.0047 \\ 

D & BERT4Rec~\cite{bert4rec} & 0.1285 & 0.3032 & 0.0475 & 0.0555 & 0.0663 & 0.1045 & 0.0043 & 0.0226 & 0.0013 & 0.0020 & 0.0020 & 0.0058 \\

\midrule

D & ReDial~\cite{redial} & 0.1705 & 0.3077 & 0.0677 & 0.0738 & 0.0925 & 0.1222 & 0.0038 & 0.0165 & 0.0012 & 0.0017 & 0.0018 & 0.0045 \\ 

D & TG-ReDial~\cite{tgredial} & 0.1679 & 0.3327 & 0.0694 & 0.0771 & 0.0924 & 0.1286 & 0.0110 & 0.0174 & 0.0048 & 0.0050 & 0.0062 & 0.0076 \\

D, E & KBRD~\cite{kbrd} & 0.1796 & 0.3421 & 0.0722 & 0.0800 & 0.0972 & 0.1333 & 0.0201 & 0.0501 & 0.0077 & 0.0090 & 0.0106 & 0.0171 \\ 

D, E, W & KGSF~\cite{kgsf} & 0.1785 & 0.3690 & 0.0705 & 0.0796 & 0.0956 & 0.1379 & 0.0215 & 0.0643 & 0.0069 & 0.0087 & 0.0103 & 0.0194 \\ 

D, E & KGConvRec~\cite{kgconvrec} & 0.1819 & 0.3587 & 0.0711 & 0.0794 & 0.0969 & 0.1358 & 0.0220 & 0.0524 & 0.0088 & 0.0102 & 0.0119 & 0.0185 \\

D, E & KECRS~\cite{kecrs} & 0.1746 & 0.3708 & 0.0654 & 0.0748 & 0.0908 & 0.1344 & 0.0234 & 0.0615 & 0.0069 & 0.0086 & 0.0107 & 0.0190 \\

\midrule

D & BERT~\cite{bert} & 0.1608 & 0.3525 & 0.0597 & 0.0688 & 0.0831 & 0.1255 & 0.0040 & 0.0194 & 0.0011 & 0.0017 & 0.0018 & 0.0050 \\

D & XLNet~\cite{xlnet} & 0.1569 & 0.3590 & 0.0583 & 0.0677 & 0.0811 & 0.1255 & 0.0040 & 0.0187 & 0.0011 & 0.0017 & 0.0017 & 0.0048 \\

D & BART~\cite{bart} & 0.1693 & 0.3783 & 0.0646 & 0.0744 & 0.0888 & 0.1350 & 0.0047 & 0.0187 & 0.0012 & 0.0017 & 0.0020 & 0.0048 \\

\midrule

D, E & \textbf{MHIM} & \textbf{0.1966*} & \textbf{0.3832*} & \textbf{0.0742*} & \textbf{0.0830*} & \textbf{0.1027*} & \textbf{0.1440*} & \textbf{0.0300*} & \textbf{0.0783*} & \textbf{0.0108*} & \textbf{0.0129*} & \textbf{0.0152*} & \textbf{0.0256*} \\

\bottomrule
\label{rec_result}
\end{tabular}

}
\end{table*}

\subsection{Experiment Setup}

\subsubsection{Datasets}

We evaluate the proposed model on \textsc{ReDial}~\cite{redial} and \textsc{TG-ReDial}~\cite{tgredial} datasets.
\textsc{ReDial} is an English conversational recommendation dataset constructed through Amazon Mechanical Turk by crowd workers under a set of comprehensive instructions.
\textsc{TG-ReDial} is a Chinese conversational recommendation dataset created semi-automatically.
The statistics of both datasets are shown in Table~\ref{datasets}.
To avoid overfitting certain user histories, we rebuild the dataset into a more strict setting by separating the data based on \texttt{user\_id} and truncating the number of historical dialogues to a certain limitation. The rebuilt dataset is also split into training, validation, and test sets in a proportion of 8:1:1. For each conversation, we start from the first sentence one by one to generate reply utterances or give recommendations by our model.
Moreover, we incorporate open-source knowledge base \textsc{DBpedia}~\cite{dbpedia} and \textsc{CN-DBpedia}~\cite{cn-dbpedia} as the external KG.

\subsubsection{Baselines}

In CRS, we consider two major tasks to evaluate the superiority of our proposed model, namely the recommendation task and the conversation task. Therefore, we compare our approach with existing CRS methods, as well as several representative recommendation and conversation models.

$\bullet$ \textbf{TextCNN}~\cite{textcnn} adopts a CNN-based model to extract personalized features from contextual utterances as user embeddings.

$\bullet$ \textbf{SASRec}~\cite{sasrec} adopts the self-attention layer to capture the dynamic patterns in user interaction sequences.

$\bullet$ \textbf{BERT4Rec}~\cite{bert4rec} adapts the original BERT~\cite{bert} model with a cloze objective loss for sequential recommendation.

$\bullet$ \textbf{Transformer}~\cite{transformer} adopts a Transformer-based encoder-decoder method to generate conversational responses.

$\bullet$ \textbf{ReDial}~\cite{redial} consists of a dialogue generation module based on HRED~\cite{hred} and a recommender module based on auto-encoder~\cite{autorec}.

$\bullet$ \textbf{TG-ReDial}~\cite{tgredial} presents the task of topic-guide conversational recommendation, and utilizes both historical interaction and dialogue text for deriving user preference in recommender module.

$\bullet$ \textbf{KBRD}~\cite{kbrd} is a knowledge-based CRS model that utilizes R-GCN to construct user representations on DBpedia, and ranks the items by dot-product for recommendations.

$\bullet$ \textbf{KGSF}~\cite{kgsf} is a knowledge-based CRS model that utilizes both word-oriented and item-oriented KGs, and aligns the two semantic spaces using Mutual Information Maximization (MIM).

$\bullet$ \textbf{KGConvRec}~\cite{kgconvrec} incorporates pre-trained entity embeddings supplemented with positional embeddings to obtain better entity representations for recommendation.

$\bullet$ \textbf{KECRS}~\cite{kecrs} proposes the Bag-of-Entity loss and the infusion loss to better integrate KGs and generate more diverse responses for recommendation.

$\bullet$ \textbf{BERT}~\cite{bert} is a language model pre-trained with the masked language model task, and we utilize the representation of the \texttt{[CLS]} token for recommendation.

$\bullet$ \textbf{XLNet}~\cite{xlnet} is a language model pre-trained using an auto-regressive method to learn bidirectional contexts, and we utilize the representation of the \texttt{[CLS]} token for recommendation.

$\bullet$ \textbf{BART}~\cite{bart} is a language model pre-trained with the denoising auto-encoding task, and we also use the representation of the \texttt{[CLS]} token for recommendation.

Among these baselines, Transformer~\cite{transformer} is the state-of-the-art text generation method,
BERT~\cite{bert}, XLNet~\cite{xlnet} and BART~\cite{bart} are pre-trained language models~(PLMs),
TextCNN~\cite{textcnn}, SASRec~\cite{sasrec} and BERT4Rec~\cite{bert4rec} are recommendation methods,
and ReDial~\cite{redial}, TG-ReDial~\cite{tgredial}, KBRD~\cite{kbrd}, KGSF~\cite{kgsf}, KGConvRec~\cite{kgconvrec}, KECRS~\cite{kecrs} are CRS methods.
Besides, we do not compare UCCR~\cite{uccr} because in their experiments, different sessions of one user are divided into train, valid, and test sets, and the user preferences are memorized by model parameters.
However, we strictly separate users based on data partition, and our method can capture user interest from historical data online.
Therefore, it is unfeasible for these two methods to achieve a fair comparison under the same data partition.

\subsubsection{Evaluation Metrics}

In our experiments, we adopt different metrics to evaluate the two tasks.
For the recommendation task, we evaluate whether our approach is able to provide item recommendations accurately. Thus, we adopt Recall@$K$, MRR@$K$, NDCG@$K$ for evaluation~($k$=10, 50).
For the conversation task, we use Distinct $n$-gram~($n$=2, 3, 4) to measure the degree of diversity for text tokens, which is calculated as the number of distinct $n$-grams scaled by the total number of sentences in the test set.

\subsubsection{Implementation Details}

We implement our approach with PyTorch\footnote{https://pytorch.org/}.
For text processing, the lengths of the current and the historical dialogue utterances are truncated to 256 and 1024, respectively.
The dimensions of embeddings are set to 300 and 128, respectively, for conversation and recommender modules.
The number of layers is set to 1 for R-GCN~\cite{r-gcn} and hypergraph convolution considering effectiveness and efficiency, and the normalization constant of R-GCN is set to 1.
The trade-off hyper-parameter $\beta$ is set to 0.9. We use Adam~\cite{adam} optimizer with the default parameter setting, and the learning rate is set to 0.001.
For recommendation, the batch size is set to 256 and 64 on \textsc{ReDial} and \textsc{TG-ReDial} respectively, and for conversation, the batch size is consistently set to 128.

For R-GCN pre-training, we generate subgraphs using the random walk API provided by DGL~\cite{dgl}, the random walk hop is set to 128 with a restart probability of 0.5. We pre-train our R-GCN for 120 steps and use Adam optimizer with learning rate of 0.005, $\beta_1 = 0.9$, $\beta_2 = 0.999$, weight decay of 1e-4, and learning rate warm-up over the first $10\%$ steps.
We use a batch size of 1024, dictionary size of 16384, temperature of 0.07, and momentum of 0.999.

\subsection{Evaluation on Recommendation Task}

In this subsection, we conduct a series of experiments to verify the effectiveness of our proposed model MHIM for the recommendation task. The results are presented in Table~\ref{rec_result}.

\begin{table}[]
\centering
\caption{
Experiment results of ablation and variation study of our model on recommendation task. We report the results of Recall@$\mathbf{\emph{K}}$, which is abbreviated as R@$\mathbf{\emph{K}}$.
}

\setlength{\tabcolsep}{2.5mm}{

\begin{tabular}{lcccc}
\toprule

\multirow{2}{*}{\textbf{Model}} & \multicolumn{2}{c}{\textsc{ReDial}} & \multicolumn{2}{c}{\textsc{TG-ReDial}} \\
\cmidrule(lr){2-3} \cmidrule(lr){4-5}
& \multicolumn{1}{c}{R@10} & \multicolumn{1}{c}{R@50} & \multicolumn{1}{c}{R@10} & \multicolumn{1}{c}{R@50} \\

\midrule

\textbf{MHIM} & 0.1966 & \textbf{0.3832} & \textbf{0.0300} & \textbf{0.0783} \\

\midrule

\text{\ \ w/o\ \ }Contrast & 0.1946 & 0.3777 & 0.0218 & 0.0605 \\

\text{\ \ w/o\ \ }Session & 0.1943 & 0.3816 & 0.0266 & 0.0713 \\

\text{\ \ w/o\ \ }Knowledge & 0.1944 & 0.3791 & 0.0286 & 0.0767 \\

\text{\ \ w/o\ \ }HyperConv & 0.1925 & 0.3823 & 0.0252 & 0.0711 \\

\text{\ \ w/o\ \ }Extension & \textbf{0.1975} & 0.3829 & 0.0287 & 0.0774 \\

\bottomrule
\end{tabular}

}
    
\label{ablation}

\end{table}

\subsubsection{Result Analysis}

Table~\ref{rec_result} shows the experiment results of different methods on recommendation task. As we can see, the CRS methods outperform recommendation methods generally~(\eg BERT4Rec). The reason might be that recommendation methods utilize item interaction records to capture user preferences, while the CRS methods further incorporate textual information from dialogues. Moreover, the CRS methods integrate the recommendation module and the conversation module seamlessly, which are mutually beneficial to each other.
For the CRS methods, we can see that the KG-enhanced methods KBRD, KGSF, KGConvRec and KECRS perform better than ReDial and TG-ReDial. This is because the KG bridges the gap between unstructured textual information and structural semantics, which promotes the user preference modeling.
For the pre-trained language models BERT, XLNet and BART, even they do not utilize external KG, they perform as well as the CRS methods. One possible reason might be that the pre-training stage on the large-scale text data learns prior knowledge which is beneficial to the downstream recommendation task.
Moreover, we notice that the KG-enhanced CRS methods outperform the other baselines by a large margin on the \textsc{TG-ReDial} dataset. One reason is that the item interactions on \textsc{TG-ReDial} are sparser than \textsc{ReDial} (Table~\ref{datasets}), and therefore, the enhancement effect of the KGs is more obvious.

Our proposed method MHIM outperforms all the baselines.
We improve the KG encoder via contrastive subgraph discrimination, and learn informative user representations via multi-grained hypergraph convolution for both recommendation and conversation tasks.
Compared to KGSF, even though we do not utilize external word-level KG, our method outperforms it significantly. Compared to PLM-based methods, our method is lighter and runs faster.

\subsubsection{Ablation Study}

In order to evaluate the effectiveness of each component, we conduct the ablation study based on different variants of MHIM, including:
(1) \textit{MHIM w/o Contrast} removes the contrastive pre-training stage (Section~\ref{sec-contrastive}) of the KG encoder;
(2) \textit{MHIM w/o Session} removes the session-based hypergraphs (Section~\ref{sec-session-based});
(3) \textit{MHIM w/o Knowledge} removes the knowledge-based hypergraphs (Section~\ref{sec-knowledge-based});
(4) \textit{MHIM w/o HyperConv} removes the hypergraph convolution (Equation~(\ref{eq-hyperconv})) on both session- and knowledge-based hypergraphs;
(5) \textit{MHIM w/o Extension} removes the hyperedge extension procedure (Section~\ref{sec-hyperedge-extension}).
Note that compared to (2) and (3), \textit{MHIM w/o HyperConv} only removes the convolution operations performed on hypergraphs, and remains the original representations of items in both session- and knowledge-based hypergraphs.
Our motivation for designing variant (4) is to validate the effectiveness of hypergraph convolution itself.

The results are shown in Table~\ref{ablation}. Firstly, we can observe that removing the pre-training stage of the KG encoder leads to the largest performance decrease, which promotes the generalization performance of GNN module.
Another observation is that both session- and knowledge-based hypergraphs and the hypergraph convolution lead to increased performance, which capture multi-grained user interest for recommendation.
Finally, we notice that for metric Recall@10, \textit{MHIM w/o Extension} performs better on \textsc{ReDial}. One possible reason is that the hypergraph extension introduces noise, which adversely affects item recommendation.

\begin{figure}
\setlength{\abovecaptionskip}{5pt}
    \centering
    \includegraphics[width=0.47\textwidth]{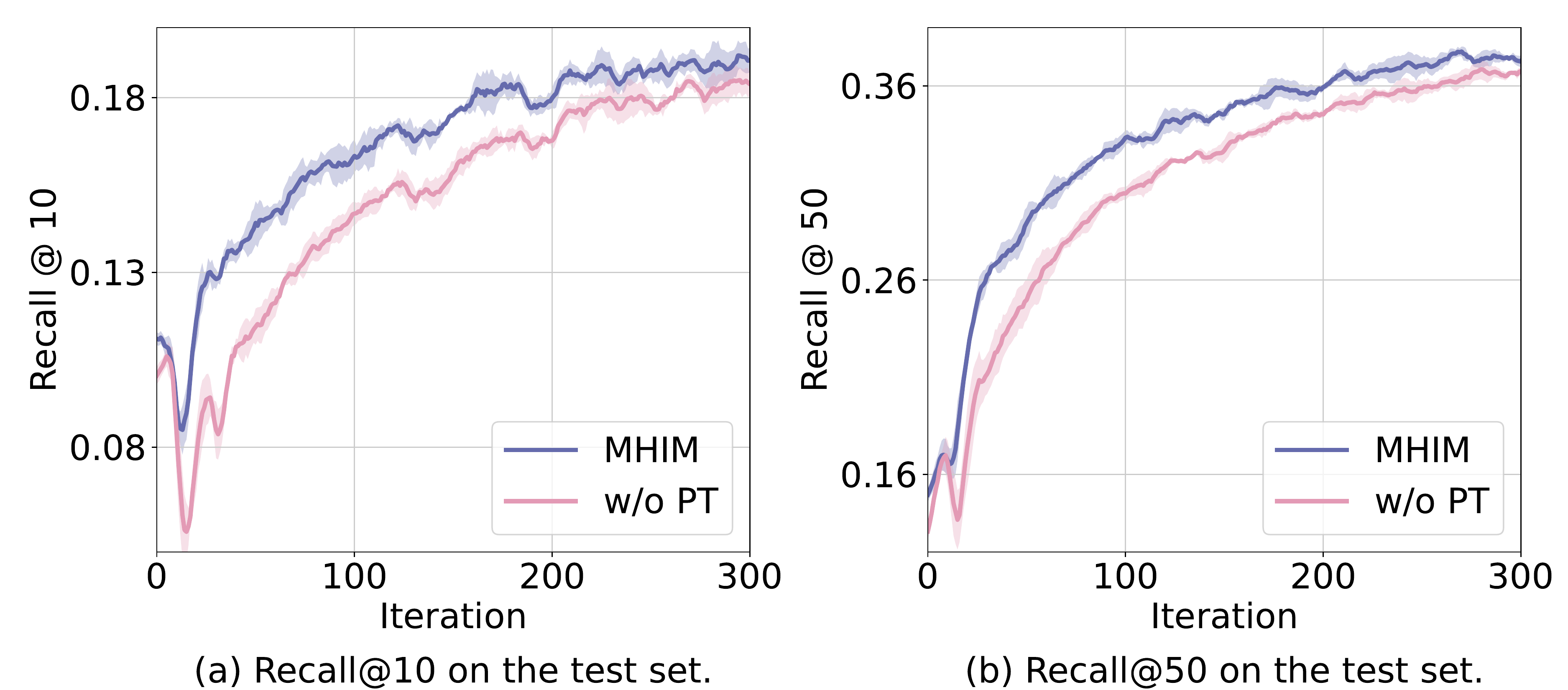}
    \caption{Performance~(\ie Recall@10 and Recall@50) comparison of MHIM and the variant \textit{without} contrastive pre-training on the \textsc{ReDial} dataset.}
    \label{effect}
\end{figure}

\subsubsection{The Effect of Pre-training Technique}

We adopt the contrastive learning technique for pre-training the KG encoder. As shown in Table~\ref{ablation}, it significantly contributes to the final performance, since it benefits multi-grained representation learning for entities. We would like to further study whether the improvement is consistent with the increase of the iteration number. Therefore, we gradually increase the iteration number on the train set, and report the corresponding evaluation metrics~(\ie Recall@10 and Recall@50) on the test set. As shown in Figure~\ref{effect}, our model can achieve an equal performance with fewer iterations compared to the variant without pre-training, and finally outperforms the latter.

\subsection{Evaluation on Conversation Task}

In this section, we verify the effectiveness of our proposed model MHIM for the conversation task. We present the results of the evaluation metrics for different methods in Table~\ref{conv_result}.

\subsubsection{Result Analysis}

As we can see, among the four baselines, the performance order is consistent with KBRD > KGSF > Transformer > ReDial. The reason is that KBRD introduces KG-based vocabulary bias to generate responses that are more consistent with user interest, and KGSF incorporates cross-attention with embeddings from entity- and word-level KGs to develop KG-enhanced decoder. However, Transformer and ReDial only utilize token sequences, ignoring user preferences hidden under the entities.
Compared with these baselines, our model MHIM consistently performs better. In our approach, the cross-attention layer and the user interest-aware bias can effectively inject the user preferences learned from multi-grained hypergraph convolution into the decoder. Therefore, our model can generate diverse responses consistent with user interest.

\begin{table}[]
\centering
\caption{Experimental results on conversation task. * indicates statistically significant improvement ($p<0.05$) over all baselines.. We abbreviate Distinct-2,3,4 as Dist-2,3,4, and ``Trans.'' refers to the Transformer model.
}

\setlength{\tabcolsep}{0.9mm}{

\begin{tabular}{lcccccc}
\toprule

\multirow{2}{*}{\textbf{Model}} & \multicolumn{3}{c}{\textsc{ReDial}} & \multicolumn{3}{c}{\textsc{TG-ReDial}} \\
\cmidrule(lr){2-4} \cmidrule(lr){5-7}
& \multicolumn{1}{c}{Dist-2} & \multicolumn{1}{c}{Dist-3} & \multicolumn{1}{c}{Dist-4} & \multicolumn{1}{c}{Dist-2} & \multicolumn{1}{c}{Dist-3} & \multicolumn{1}{c}{Dist-4} \\

\midrule

ReDial & 0.0214 & 0.0659 & 0.1333 & 0.2178 & 0.5136 & 0.7960 \\
Trans. & 0.0538 & 0.1574 & 0.2696 & 0.2362 & 0.7063 & 1.1800 \\
KBRD & 0.0765 & 0.3344 & 0.6100 & 0.8013 & 1.7840 & 2.5977 \\
KGSF & 0.0572 & 0.2483 & 0.4349 & 0.3891 & 0.8868 & 1.3337 \\

\midrule

\textbf{MHIM} & \textbf{0.3278*} & \textbf{0.6204*} & \textbf{0.9629*} & \textbf{1.1100*} & \textbf{2.3520*} & \textbf{3.8200*} \\

\bottomrule
\end{tabular}

}
    
\label{conv_result}

\end{table}

\begin{figure}
\setlength{\abovecaptionskip}{5pt}
    \centering
    \includegraphics[width=0.47\textwidth]{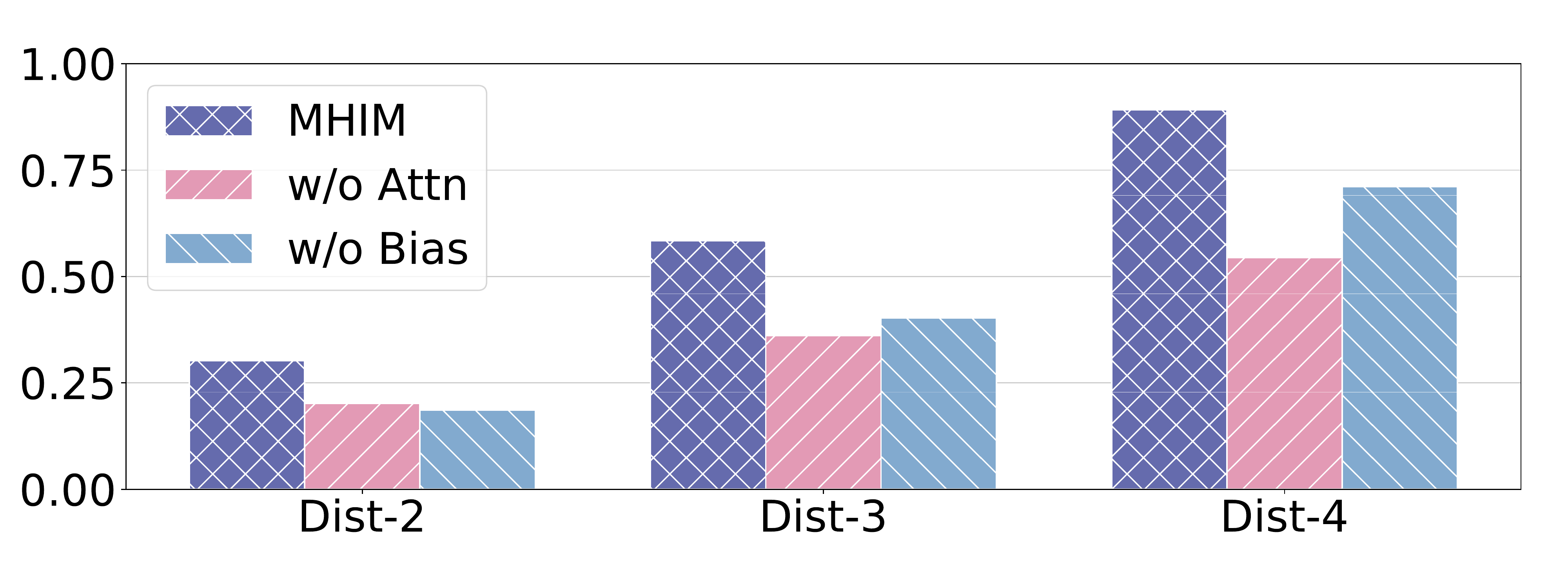}
    \caption{
    Experiment results of Distinct-2,3,4, abbreviated as Dist-2,3,4, on the \textsc{ReDial} dataset about the conversation task, and the ablation study shows that MHIM consistently outperforms its variants \textit{w/o Attn} and \textit{w/o Bias}.
    }
    \label{conv-ablation}
\end{figure}

\subsubsection{Ablation Study}

We also conduct ablation study to evaluate the effectiveness of each component for conversation task on the \textsc{ReDial} dataset. We devise two variants of our model:
(1) \textit{MHIM w/o Attn} removes the cross-attention layer considering item representations enhanced by multi-grained hypergraph convolution (Equation~(\ref{eq-mha-nsk}));
(2) \textit{MHIM w/o Bias} removes the user interest-aware bias for token prediction (Equation~(\ref{eq-bias})).
We report the results of Distinct-2,3,4 for evaluation.

The results are shown in Figure~\ref{conv-ablation}. We can observe that removing any component leads to performance decrease, indicating that both components are beneficial for user preference modeling and contribute to more diverse responses. In addition, removing the cross-attention layer leads to larger performance decrease, which evaluates the effectiveness of the proposed multi-grained hypergraph convolution for item recommendation.

\subsection{Contrastive Pre-training Settings}

As introduced above, the contrastive pre-training procedure for R-GCN encoder is applied on the large-scale, extended KG. However, if memory is allocated to each node, the storage and time costs will be unacceptable.
Fortunately, some critical nodes are more likely to appear in random walk sequences, potentially connecting to more edges in the KG.
As a result, we only allocate memory space to critical nodes that appear frequently, implying that the other nodes are associated with a specific \texttt{\_\_UNKNOWN\_\_} embedding.

\begin{figure}
    \centering
    \includegraphics[width=0.47\textwidth]{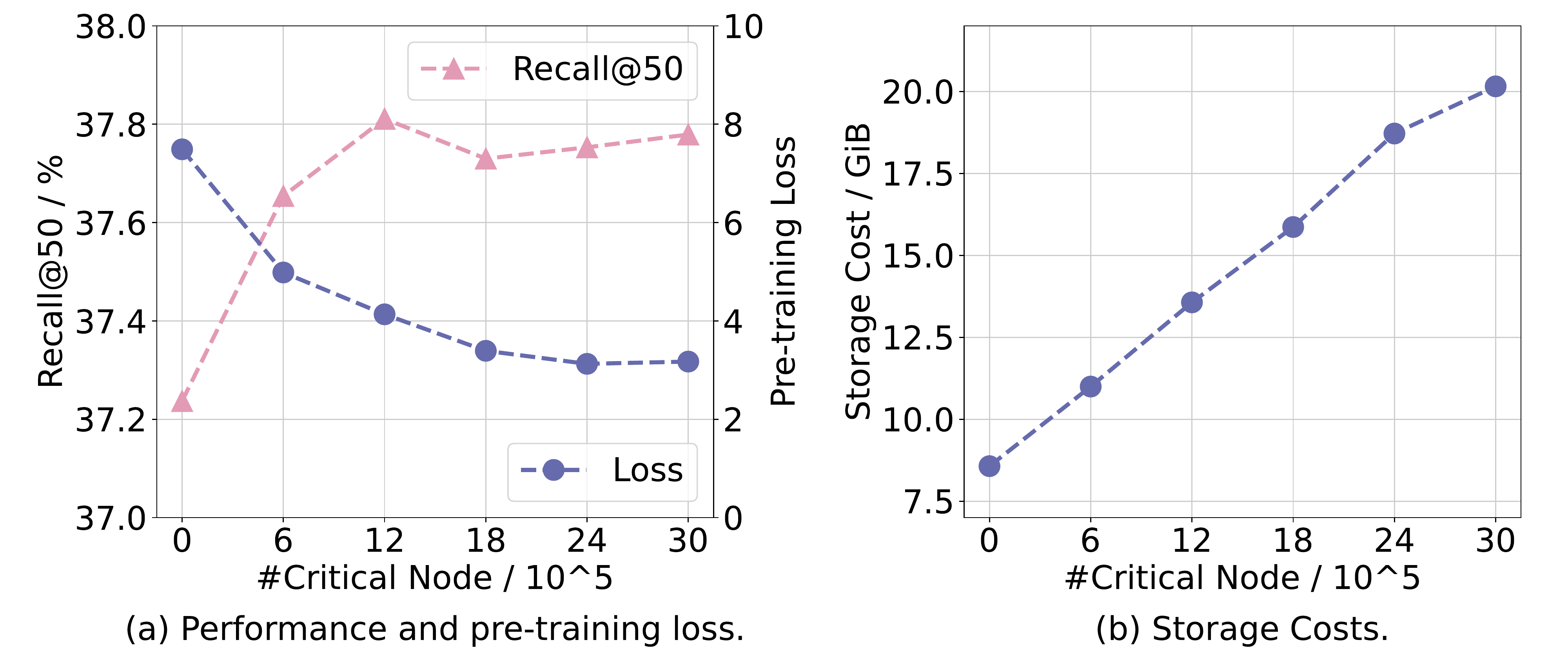}
    \caption{The recommendation performances, pre-training loss and storage costs with different numbers of critical nodes on the \textsc{ReDial} dataset.}
    \label{nodenum}
\end{figure}

To determine the number of critical nodes, we conduct a series of experiments based on different critical node numbers from 0 to 3,000,000 on \textsc{DBpedia}. As we can see in Figure~\ref{nodenum}, while the node number is set to 600,000, the pre-training loss and the Recall@50 in the recommendation task on \textsc{ReDial} both reach the inflection point, and furthermore, the storage cost is slightly higher than the group whose critical nodes number is 0. Therefore, we retain 600,000 critical nodes on \textsc{DBpedia} in our experiments. Similarly, we retain 1,540,000 critical nodes on \textsc{CN-DBpedia} for \textsc{TG-ReDial}.

\section{RELATED WORK}

In the following section, we first introduce the prior work on Conversational Recommender System~(CRS) and session-based recommendation. Then, we briefly review the existing literature on graph representation learning, including pre-training on graph neural networks and hypergraph learning.

\subsection{Recommender System}

\paratitle{Conversational Recommendation}. Conversational recommender systems model user interest through multi-turn dialogues and provide high-quality recommendations. Existing studies about CRS can be roughly divided into two categories, \ie attribute-based CRS and generation-based CRS.

Attribute-based CRS~\cite{crm, saur, ear, cpr, unicorn, kbqg, fpan} typically captures user preferences by asking queries about item attributes and generating responses using pre-defined templates~\cite{crm, ear}. Most of these methods gradually narrow down the hypothesis space to search for the proper items within fewer turns. However, this kind of CRS does not pay enough attention to generating human-like responses in natural language, which may hurt user experiences.

Generation-based CRSs~\cite{redial, kbrd, kgsf, dcr, revcore, ntrd, uccr, c2crs, unimind, mese, barcor, kgconvrec, kecrs} alleviate this problem by adopting the Seq2Seq architecture~\cite{hred, transformer} to generate fluent utterances as responses, which constructs an end-to-end framework for both conversation task and recommendation task. Researchers release a benchmark dataset \textsc{ReDial}~\cite{redial} which contains human conversations about movie recommendation. Further studies incorporate external data to improve user preference modeling and recommendation, including entity-oriented knowledge graph~\cite{kbrd, kecrs, kgconvrec}, word-oriented knowledge graph~\cite{kgsf} and review information~\cite{revcore}. To effectively leverage external data, researchers propose a coarse-to-fine contrastive learning framework~\cite{c2crs} to improve data semantic fusion. A more recent study~\cite{uccr} first highlights that the user's historical dialogue sessions and look-alike users are essential for user preference modeling.

However, the user interest that lies beneath complicated historical data has yet to be comprehensively captured.
Our work extends the second category of research by leveraging historical dialogue sessions and large-scale external knowledge. The key novelty lies in the user interest modeling through multi-grained hypergraph convolution, which can effectively model historical user interest for better recommendation.

\paratitle{Session-based Recommendation}. Session-based recommendation focuses on capturing dynamic user interests for recommendation based on short-term sessions, where a session refers to multiple user-item interactions that happen over a short period of time. GRU4Rec~\cite{gru4rec} utilizes gated recurrent units~(GRUs) to model user behaviors sequentially, which helps to utilize complex intra- and inter-session relations for recommendation. NARM~\cite{narm} proposes to incorporate an attention mechanism into recurrent neural networks~(RNNs) to capture user interests accurately.
Recently, graph neural networks~(GNNs) have been adopted for session-based recommendation~\cite{wu2019session, xu2019graph, song2019session}, because of their great potential in modeling complex graph-structured context data.
Moreover, researchers propose to extend the session-based scenarios to multi-session-based scenarios~\cite{wang2022effectively}, integrating more context information for accurate recommendation.
Compared with session-based recommendation, our work considers a conversation user takes part in as a session, and captures user interest by modeling short-term sequences for conversational recommendation.

\subsection{Graph Representation Learning}

\paratitle{Graph Neural Network}. Due to their tremendous capacity to model graph-structured data, Graph Neural Networks (GNNs) have gained a lot of attention in recent years. Existing GNN methods can be divided into spectral methods~\cite{gcn} and spatial methods~\cite{gat, graphsage, r-gcn}. Most methods follow a message passing~\cite{mpnn} scheme to aggregate structural information from nodes’ neighbors.
Though GNNs are effective for modeling graph data, they usually require abundant task-specific data for end-to-end training. 
Motivated by the recent advances in pre-training from natural language processing~\cite{bert, gpt-3} and computer vision~\cite{simclr, mae}, researchers devote efforts to pre-training on GNNs~\cite{gnn-pretrain, gpt-gnn, dgi, gcc, graphcl}. The key idea is to pre-train an expressive GNN encoder on massive unlabeled graph datasets or coarse-grained supervised datasets.
Existing works mainly focus on designing proper pre-training tasks, \eg attribute prediction~\cite{gnn-pretrain}, graph property prediction~\cite{gnn-pretrain}, graph reconstruction~\cite{gpt-gnn} and contrastive learning~\cite{dgi,graphcl,gcc}.
In the field of CRS, the training data remains insufficient.
Therefore, we propose to pre-train the graph encoder on large-scale KGs~\cite{dbpedia, cn-dbpedia} via contrastive learning~\cite{gcc}.

\paratitle{Hypergraph Learning}. Hypergraph~\cite{hypergraph-theory} generalizes the concept of edge to make it connect more than two nodes, and provides a natural way to capture high-order relations. It has been explored by combining with promising deep learning techniques. HGNN~\cite{hgnn} and HyperGCN~\cite{hypergcn} are the first to design hypergraph convolution operations to handle high-order correlations. The attention mechanism is further introduced to improve the performance~\cite{hgnn-attn}. There are also several studies combining hypergraph learning with recommender systems~\cite{hyperrec, dhcf, dhcn, mhcn, xia2022hypergraph}. HyperRec~\cite{hyperrec} uses hypergraph to model the short-term user preference for next-item recommendation. DHCF~\cite{dhcf} captures high-order correlations among users and items for general collaborative filtering. DHCN~\cite{dhcn} further exploits inter-hyperedge information for session-based recommendation. MHCN~\cite{mhcn} integrates self-supervised learning into the training of the hypergraph convolutional network for social recommendation. Our work is the first to model user interest with hypergraph learning for conversational recommendation.
\section{CONCLUSION AND FUTURE WORK}

In this paper, we propose a novel multi-grained hypergraph interest modeling framework to model user interest for conversational recommendation. By employing hypergraphs to model historical dialogue sessions and reconstruct external KGs, we obtain session- and knowledge-based hypergraphs, which help to comprehensively capture user interest lies beneath complicated historical conversations from multi-grained perspectives. We then pre-train the KG encoder with the subgraph instance discrimination task, and learn item representations by the multi-grained hypergraph convolution. 
Finally, the proposed user interest-aware CRS is able to provide proper item recommendation and give informative responses.
Extensive experiments on two datasets show that our approach yields better performance than several competitive baselines.

For future work, we consider designing a unified user interest learner incorporating multi-type historical data. Besides, it is also interesting to devise an interest transfer module, through which the learned preferences will benefit each other among different users.

\section*{Acknowledgement}

This work was partially supported by National Natural Science Foundation of China under Grant No. 62222215, Beijing Natural Science Foundation under Grant No. 4222027, and Beijing Outstanding Young Scientist Program under Grant No. BJJWZYJH012019100020098.

\bibliographystyle{ACM-Reference-Format}
\bibliography{main}

\end{document}